\documentclass[preprint,showpacs,preprintnumbers,superscriptaddress,amsmath,amssymb]{revtex4}

\usepackage{amssymb}
\usepackage[dvips]{graphicx}
\usepackage{dcolumn}
\usepackage{bm}
\usepackage{subfigure}
\usepackage{floatflt}
\usepackage{float}
\usepackage{rotating}
\usepackage{multirow}
\usepackage[bookmarksnumbered,bookmarksopen,colorlinks,citecolor=blue,linkcolor=blue]{hyperref} %

\begin{document}

\title{Nuclear symmetry energy at subnormal densities from measured nuclear masses}

\author{Min Liu }%
 \email{lium$_$816@hotmail.com}
 \affiliation{Key Laboratory of Beam Technology and Material Modification of Ministry of Education,
              College of Nuclear Science and Technology, Beijing Normal University, Beijing 100875, China}
 \affiliation{Department of Physics, Guangxi Normal University, Guilin 541004, China}
 \affiliation{Beijing Radiation Center, Beijing 100875, China}

\author{Ning Wang}%
 \affiliation{Department of Physics, Guangxi Normal University, Guilin 541004, China}

\author{Zhuxia Li}%
 \email{lizwux@ciae.ac.cn}
 \affiliation{ China Institute of Atomic Energy, Beijing 102413, China}

\author{Fengshou Zhang}%
 \email{fszhang@bnu.edu.cn}
 \affiliation{Key Laboratory of Beam Technology and Material Modification of Ministry of Education,
              College of Nuclear Science and Technology, Beijing Normal University, Beijing 100875, China}
 \affiliation{Beijing Radiation Center, Beijing 100875, China}

\begin{abstract}
The symmetry energy coefficients for nuclei with mass number
$A=20\sim 250$ are extracted from more than 2000 measured nuclear
masses. With the semi-empirical connection between the symmetry
energy coefficients of finite nuclei and the nuclear symmetry
energy at reference densities, we investigate the density
dependence of symmetry energy of nuclear matter at subnormal
densities. The obtained results are compared with those extracted
from other methods.
\end{abstract}


\maketitle

\begin{center}
\textbf{I. INTRODUCTION}
\end{center}

The nuclear symmetry energy  $e_{\rm sym}(\rho)$, which represents
the energy cost per nucleon to convert all the protons to neutrons
in symmetric nuclear matter, has attracted lots of attention in
astrophysics and nuclear physics, because it intimately relates to
a wealth of astrophysical phenomena, the structure character of
nuclei and the dynamical process of nuclear reactions. The density
dependence is a key point in the study of symmetry energy. Many
theoretical and experimental efforts have been paid to constrain
the density dependence of the symmetry energy
\cite{Tsang09,Zhang08,Li08}. Presently, some constraints on the
symmetry energy at subnormal densities have been made from the
double n/p ratio and isospin diffusion in intermediate energy
heavy ion collisions of isospin asymmetric nuclei
\cite{Tsang09,Zhang08,Li05} and from the nuclear properties such
as the thickness of neutron skin and the binding energy of finite
nuclei \cite{Cent09,Klim07,Myers96,Dani03}. The uncertainty of the
symmetry energy coefficient and the density dependence of symmetry
energy at subnormal densities is still large, and more study is
still needed.

We try in this work to constrain the symmetry energy from more
than 2000 precisely measured nuclear masses. By directly fitting
the measured nuclear masses with the liquid drop mass formula, one
can obtain the symmetry energy coefficients of nuclei in which
both volume and surface term are included
\cite{Dani09,Kirson,Wang10b}. It is known that the symmetry energy
coefficients of finite nuclei $a_{\rm sym}$ are considerably
smaller than that of the infinite nuclear matter due to the
influence of the surface region of nuclei. A semi-empirical
connection between the symmetry energy of nuclear matter at
reference density and the properties of finite nuclei was proposed
in \cite{Bald04}. More recently, a relation $a_{\rm sym}(A) =
e_{\rm sym}(\rho_A)$ between the symmetry energy coefficients of
finite nuclei and the symmetry energy of nuclear matter at
reference density was proposed in \cite{Cent09}. For $^{208}$Pb,
the reference density has a value of about $\rho_{208} \simeq 0.1
$ fm$^{-3}$. The relation provides a possible way to explore the
property of nuclear matter from the property of finite nuclei.
Combining this relation and the symmetry energy coefficients of
finite nuclei extracted from the measured nuclear masses, we
investigate the density dependence of nuclear symmetry energy at
subnormal densities.

The paper is organized as follows: In Sec.II, we determine the
symmetry energy coefficients for nuclei with mass number $A=20
\sim 250$ by analyzing the measured nuclear masses \cite{Audi03}.
In Sec.III, we constrain the nuclear symmetry energy at subnormal
density with the relation of $a_{\rm sym}(A) = e_{\rm
sym}(\rho_A)$, and compare our results with those obtained by
using other approaches. The reference density as a function of
nuclear mass number is also deduced. A short summary is given in
Sec.IV.

\begin{center}
\textbf{II. SYMMETRY ENERGY COEFFICIENTS OF FINITE NUCLEI}
\end{center}

In Ref.\cite{Wang10a}, the symmetry energy coefficients of finite
nuclei were studied. The energy per particle $e(A,I)$ of a nucleus
can be expressed as a function of mass number $A$ and
isospin-asymmetry $I=(N-Z)/A$ according to the Weizs\"acker
nuclear energy formula:
\begin{eqnarray}
  e(A,I)&=& a_{\rm v} + a_{\rm s} A^{-1/3} + e_{\rm Coul}(A,I) + a_{\rm sym}(A)I
  ^2  + a_{\rm p} A^{-3/2}\Delta_{\rm np} + e_{\rm w},
\end{eqnarray}
with
\begin{eqnarray}
\Delta_{\rm np} = \begin{cases} 1& \text{for even-even nuclei},\\
0& \text{for odd-$A$ nuclei},\\-1& \text{for odd-odd nuclei}.
\end{cases}
\end{eqnarray}
Here a small correction term, i.e., the Wigner term $e_{\rm w}$ is
introduced for a better description of the systematic behavior in
the symmetry energy coefficients of nuclei. The $a_{\rm v}$,
$a_{\rm s}$ and $a_{\rm p}$ denote the coefficients of the volume,
the surface, and the paring term, respectively. Subtracting the
Coulomb term and the Wigner term from the energy per particle, one
obtains
\begin{eqnarray}
  e_{\rm m}(A,I) &=&   e(A,I) - e_{\rm Coul}(A,I) - e_{\rm w}    \nonumber \\
     &=& e_0(A)+ a_{\rm sym}(A)I ^2.
\end{eqnarray}
Here we assume that the $a_{\rm v}$ and $a_{\rm s}$ terms are
independent of isospin asymmetry and can be rewritten as $e_0(A)$.
For the Coulomb energy, we take the same form as in Ref.
\cite{Wang10a}. For the Wigner term $e_{\rm w}=E_{\rm w}/A$, we
take the form as in \cite{Myers97}
\begin{eqnarray}
 E_{\rm w} = -C_0 \exp(-W|I|/C_0)
\end{eqnarray}
with two parameters $C_{0}=-10$ MeV and $W=42$ MeV, which is a
direct consequence of the independent-particle model. We have
checked that the obtained symmetry energy coefficients of finite
nuclei do not change appreciately  by varying the parameter $W$ in
a reasonable region of $42\sim47$ MeV \cite{Sat}.

\begin{figure}
\includegraphics[angle=0,width=1 \textwidth]{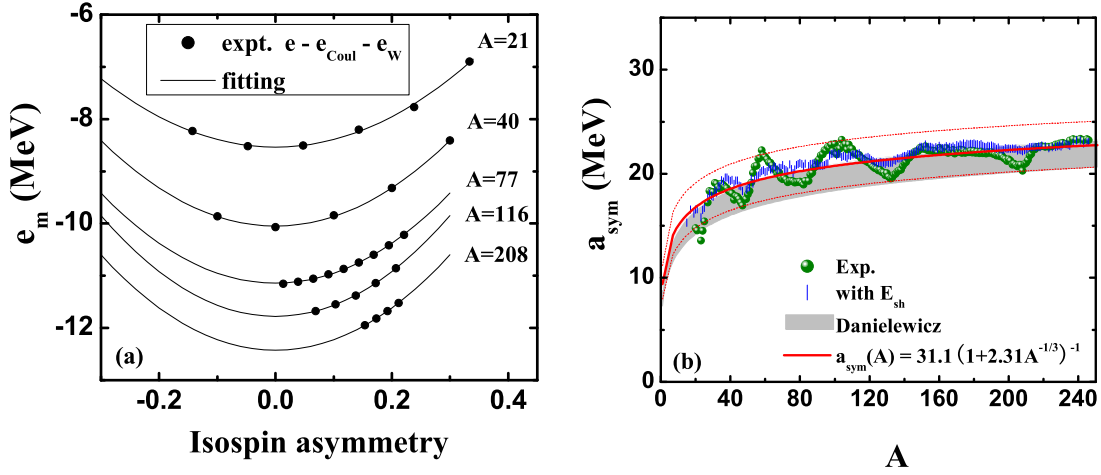}
\caption{(Color online) (a) Value of $e_{\rm m}$ as a function of
isospin asymmetry for selected series of isobaric nuclei. The
filled circles and the solid curves denote the experimental data
and  the results by fitting the experimental data, respectively.
(b) Symmetry energy coefficients of nuclei as a function of mass
number. The filled circles denote the extracted results from the
measured nuclear masses. The shades denote the results of
Danielewicz et al \cite{Dani09}. The vertical short dashes denote
the results when the shell corrections of nuclei from Ref.
\cite{Wang10b} are removed from measured energy per particle. The
red curves denote the results by fitting the circles.}
\end{figure}

Now we can take $e_0(A)$ and $a_{\rm sym}(A)$ as parameters and
perform a two-parameter parabola fitting to the $ e_m(A,I)$ for
each series nuclei with the same mass number A. The experimental
$e(A,I)$ data are taken from the mass table AME2003 \cite{Audi03}.
For nuclei with even mass number, only even-even nuclei are taken
into account in our calculations to consider the pairing effects.
All available isobaric nuclei with mass number $A = 20 \sim 250$
are considered in the calculations. Fig.1 (a) shows the values of
$e_{\rm m}$ as a function of isospin asymmetry for series of
isobaric nuclei with $A=21, 40, 77,116,208$, as examples. The
filled circles and the solid curves denote the experimental data
and the results with two-parameter parabola fitting, respectively.
The curvature of each curve gives the corresponding symmetry
energy coefficient $a_{\rm sym} (A)$ of the nuclei with mass
number $A$. The extracted symmetry energy coefficients of finite
nuclei as a function of mass number are shown in Fig.1(b). The
filled circles denote the extracted $a_{\rm sym} (A)$ from the
measured nuclear masses. The shades denote the results of
Danielewicz et al \cite{Dani09}. The vertical short dashes denote
the results when the shell corrections of nuclei \cite{Wang10b}
are removed from the measured values of energy per particle. In
the region $A < 120$, the $a_{\rm sym}(A)$ obtained in our
approach show some oscillations and fluctuations. For heavy
nuclei, our results of $a_{\rm sym}(A)$ are comparable with those
of Danielewicz et al. When the shell corrections are taken into
account, the fluctuations in the extracted $a_{\rm sym}(A)$ are
reduced effectively. The mass dependence of the symmetry energy
coefficients of nuclei is written by Danielewicz et al.
\cite{Dani09} as
\begin{eqnarray}
a_{\rm sym}(A) = S_{0}(1 + \kappa A^{-1/3})^{-1},
\end{eqnarray}
where $S_{0}$ is the volume symmetry energy coefficient of nuclei,
i.e. the nuclear symmetry energy at normal density $e_{\rm
sym}(\rho_0)$,  and $\kappa$ is the ratio of the surface symmetry
coefficient to the volume symmetry coefficient. By performing a
two-parameter fitting to the $a_{\rm sym}(A)$ obtained previously,
i.e. the filled circles in Fig.1(b), we can obtain the values of
$S_0$ and $\kappa$. With $95\%$ confidence intervals of $S_0$ and
$\kappa$, we obtain $S_0 = 31.1 \pm 1.7$ MeV and $\kappa = 2.31
\pm 0.38 $, respectively. The results by fitting the circles are
shown in Fig.1(b) with the red curves. The obtained value of $S_0$
is in good agreement with the range of $S_0 = 30.2 \sim 33.8$ MeV
given by the pygmy dipole resonance (PDR) data \cite{Klim07}.

\begin{center}
\textbf{III. NUCLEAR SYMMETRY ENERGY AT SUBNORMAL DENSITY}
\end{center}

Now let us turn to study the density dependence of nuclear
symmetry energy at subnormal densities. In experiment, recent
research in intermediate-energy heavy-ion collisions (HIC) is
consistent with a dependence at subnormal densities
\cite{Tsang09,Li05,Shet07}
\begin{eqnarray}
  e_{\rm sym}(\rho) = S_0 (\rho/\rho_0)^{\gamma}.
\end{eqnarray}
With the relation
\begin{eqnarray}
e_{\rm sym}(\rho_A) = a_{\rm sym}(A),
\end{eqnarray}
proposed by Centelles et al. \cite{Cent09}, one can obtain the
symmetry energy for nuclear matter from the symmetry energy
coefficients of finite nuclei. The $\rho_A$ is the reference
density in nuclear matter to make the equation hold
\cite{Bald04,Cent09}, which is significantly smaller than the
saturation density because of the surface region of nuclear
density profile. It is known that the relation of Eq.(7) is hold
at $\rho_A \simeq 0.1 $ fm$^{-3}$ for $^{208}$Pb from various
effective interactions \cite{Trip08,Cent09}. Now inserting Eq.(5),
(6) into Eq.(7), one obtain
\begin{eqnarray}
(\rho_{\rm A}/\rho_0)^{\gamma} = (1 + \kappa A^{-1/3})^{-1}.
\end{eqnarray}
Applying $\rho_{\rm 208} = 0.1$  fm$^{-3}$
\cite{Cent09,Trip08,Bald04} and the confidence interval $\kappa =
2.31 \pm 0.38 $ determined by $a_{\rm sym}(A)$ in the previous
section, we obtain the range of $\gamma$
\begin{eqnarray}
\gamma = 0.7 \pm 0.1.
\end{eqnarray}

With the relation of Eq.(8), one can obtain the expression of the
reference density $\rho_{\rm A}$ for a nucleus with mass A, i.e.
\begin{eqnarray}
\rho_{\rm A} = \rho_0/(1 + \kappa A^{-1/3})^{1/\gamma}.
\end{eqnarray}
\begin{figure}
\includegraphics[angle=0,width=0.7  \textwidth]{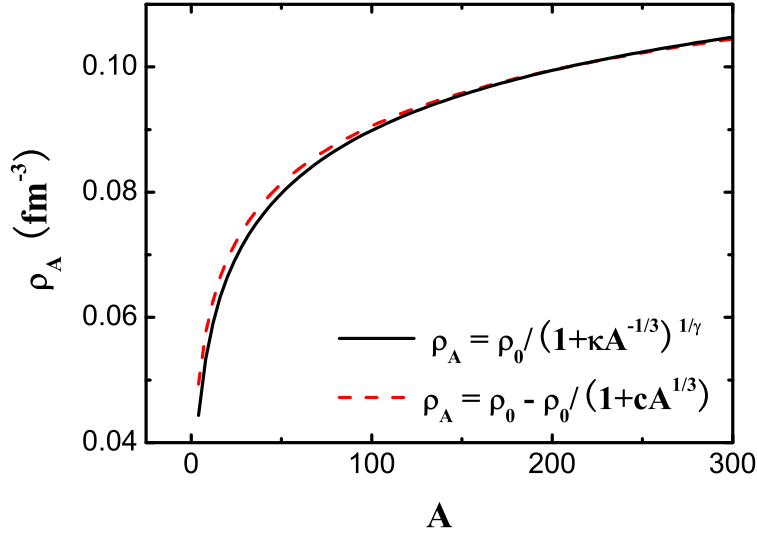}
\caption{(Color online) Reference density $\rho_A$ as a function
of mass number. The solid and the dashed curve denote the results
in this work and those from Ref. \cite{Cent09}, respectively. }
\end{figure}
The value of $\rho_A$ decreases with the decrease of nuclear size
because of the enhanced surface effects in light nuclei. Moreover,
Eq.(10) indicates that the reference density $\rho_{\rm A}$
depends on the ratio $\kappa$ and the parameter $\gamma$ which
describes the stiffness of symmetry energy. Fig.2 shows the
reference densities $\rho_A$ as a function of nuclear mass number.
The black solid curve denotes the results of Eq.(10) with $\rho_0
= 0.16$ fm$^{-3}$, $\kappa = 2.31$ and $\gamma = 0.7$. From the
figure one sees that the reference densities for finite nuclei
with $A=20 \sim 250$ cover the densities in the range of $0.42
\rho_0 \leq \rho \leq 0.64 \rho_0$. The results from the
parameterized expression of $\rho_A = \rho_0- \rho_0/(1+cA^{1/3})$
proposed by Centelles et al. in \cite{Cent09} with various
effective interactions, are also shown in the figure with the red
dashed curve for comparison. One can see that the reference
densities obtained from the two different approaches are in good
agreement with each other. The validity of the parameterized
expression proposed by Centelles et al. has been tested in the
mass region $40 \leq A \leq 208$ \cite{Cent09}. For mass region
$A<40$ and $A>208$, the two models give similar extrapolation.
From Fig.2 one sees that for heavy nuclei the reference densities
change slowly with the mass number and are close to 0.1 $\rm
fm^{-3}$, while for light nuclei the reference densities fall very
fast with decrease of mass. Therefore it is better to apply the
relation (7) by choosing the symmetry energy coefficient of heavy
nuclei to relate the symmetry energy for nuclear matter at the
reference density.

\begin{figure}
\includegraphics[width=12cm]{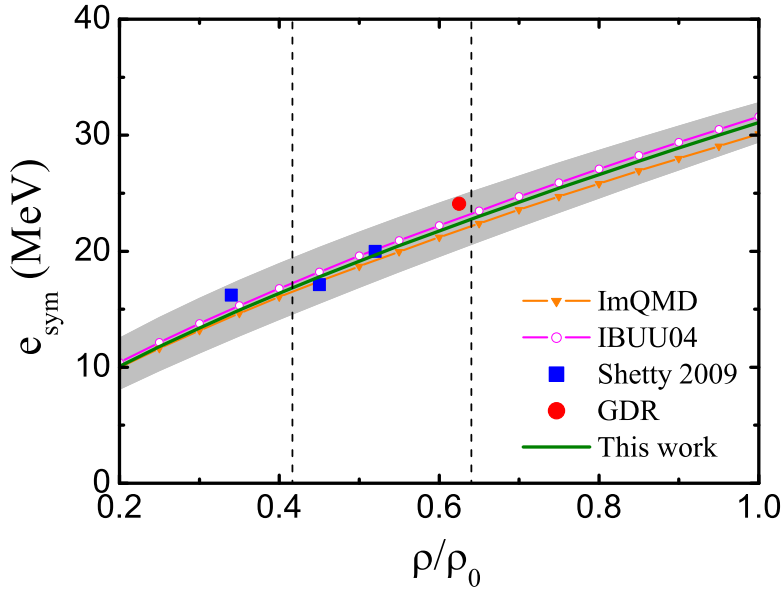}
\caption{(Color online) Density dependence of the nuclear symmetry
energy with different methods. The two vertical dashed lines give
the corresponding density region of nuclei with $A = 20 \sim 250$.
The gray shades denote the region with $S_0 = 31.1 \pm 1.7$ MeV
and $\gamma = 0.7 \pm 0.1$, using the form $e_{\rm sym}(\rho) =
S_0 (\rho/\rho_0)^{\gamma}$. }
\end{figure}

Fig.3 shows the symmetry energy of nuclear matter as a function of
$\rho/\rho_0$ obtained in this work and the comparison with the
results obtained by other methods. The green solid curve
corresponds to the symmetry energy calculated with $e_{\rm
sym}(\rho) = 31.1 (\rho/\rho_0)^{0.7}$, in which $S_0=31.1$ MeV is
the favorite value from $a_{\rm sym}(A)$ and $\gamma=0.7$ is
obtained from $\rho_{\rm A}$. The two vertical dashed lines show
the corresponding density region $0.42 \rho_0 \leq \rho \leq 0.64
\rho_0$ of nuclei with $A = 20 \sim 250$. One should note that the
results in this work for density regions ($\rho < 0.5 \rho_0$ and
$\rho > 0.63 \rho_0$, corresponding to $A<40$ and $A>208$,
respectively) represent an extrapolation. The orange dashed curve
denotes the symmetry energy constrained by comparing the
measurements of the isospin diffusion and the neutron to proton
double ratio in $^{124,112}\rm Sn + ^{124,112}\rm Sn$ reactions
with the calculations of improved quantum molecular dynamics model
(ImQMD) where $e_{\rm sym}(\rho) = 12.5(\rho/\rho_0)^{2/3} +
17.6(\rho/\rho_0)^{\gamma}$ with $\gamma=0.7$ \cite{Tsang09}. The
dot-dashed curve denotes the symmetry energy as $e_{\rm sym}(\rho)
= 31.6(\rho/\rho_0)^{\gamma}$ with $\gamma = 0.69$, obtained by
the comparison between NSCL-MSU isospin diffusion data and the
IBUU04 \cite{Li05}. The blue solid squares give the mapped $e_{\rm
sym}(\rho)$ from the correlation between temperature, excitation
energy, density and the isoscaling parameter \cite{Shet09,Shet10}.
The red filled circle gives the constrained $e_{\rm sym}(0.1)$
from the excitation energy of giant dipole resonance (GDR) in
$^{208}\rm Pb$ \cite{Trip08}. The gray shades in Fig.3 are bounded
by $S_0 = 31.1 \pm 1.7 $ MeV and $\gamma = 0.7 \pm 0.1$. One can
see from Fig.3 that the obtained nuclear symmetry energy in this
work locate between the results of ImQMD and IBUU04. The gray area
is consistent with the results from ImQMD, IBUU04, GDR and the
mapped data from isoscaling parameter in \cite{Shet09,Shet10} at
subnormal densities.

\begin{figure}
\includegraphics[width=16cm]{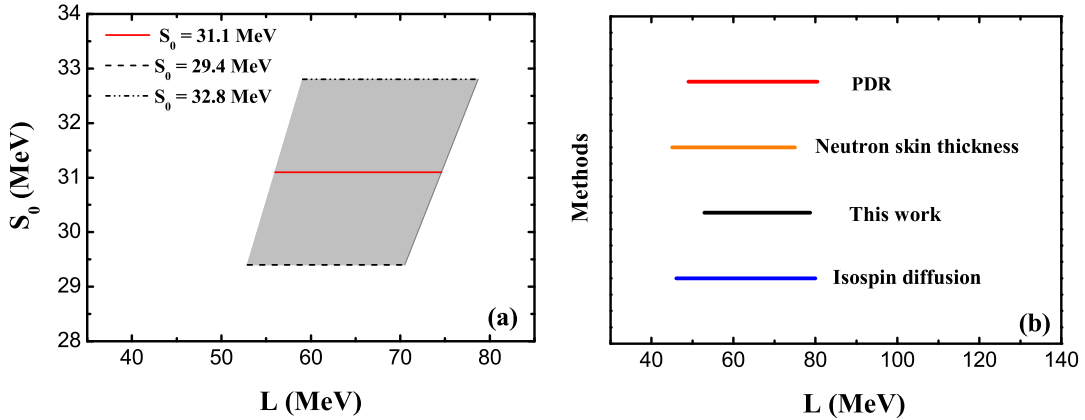}
\caption{(Color online) (a) Range of slope parameter $L$
determined by $S_0 = 31.1 \pm 1.7$ MeV and $\gamma = 0.7 \pm 0.1$
in this work. (b) Range of slope parameter $L$ determined from
different observables. The results are taken from \cite{Carb10}
with PDR, from \cite{Warda09} with neutron skin thickness and from
\cite{Tsang09} with isospin diffusion, respectively.}
\end{figure}

Based on the extracted value of $S_0$ and the extrapolated nuclear
symmetry energies at densities around the normal density, we
further study the slope parameter $L = 3\rho_0 \frac{\partial
e_{\rm sym}(\rho)}{\partial \rho}|_{\rho_0}$, which is an
effective quantity to characterize the density dependence of
symmetry energy. From previously obtained range of $\gamma = 0.7
\pm 0.1$ and the value of $S_0$, the value of $L$ can be obtained
directly. The gray shades in Fig.4(a) show the area of $L$
determined by $S_0 = 31.1 \pm 1.7$ MeV and $\gamma = 0.7 \pm 0.1$.
The range of $L$ for $S_0 = 31.1$ MeV is shown with the solid
line. The dashed and the dashed-dot-dot line correspond to the
case of $S_0 = 29.4 $ and $32.8$ MeV, respectively, which are the
boundary of the confidence interval of $S_0$. The values of $L$
for the three cases are listed in Table I. From the calculations,
we obtain the largest range of the slope parameter, $53 \lesssim L
\lesssim 79$ MeV for $S_0 = 31.1 \pm 1.7$ MeV. To compare with the
 values of $L$ obtained from other methods, we show  the
range of slope parameter $L$ determined in this work and those
from other recent analyzing with different observables in
Fig.4(b). The black solid line is the result from the measured
nuclear masses in this work. The result from isospin diffusion is
obtained from \cite{Tsang09} with $S_0 = 30.1$ MeV and $\gamma =
0.4 \sim 1.05$. The result from PDR is taken from \cite{Carb10}
with $L = 64.8 \pm 15.7$ MeV and $S_0 = 32.3 \pm 1.3 $ MeV. The
result from neutron skin thickness is taken from \cite{Warda09}.
The range of the values of $L$ obtained in this work is in
consistent with those obtained from other analyzing.
\begin{table}
\caption{Range of slope parameter $L$ for different cases of
$S_0$.}
\begin{tabular}{cccc}
 \hline\hline
        $S_0 (\rm MeV)$& $L_{\rm min} (\rm MeV)$&$L_{\rm max} (\rm MeV)$&$L (\rm MeV)$ \\
 \hline
29.4   & 52.9 & 70.6 &   $61.7 \pm 8.8$ \\
31.1   & 56 & 74.6 &   $65.3 \pm 9.3$ \\
32.8   & 59.1 & 78.7 &  $68.9 \pm 9.8$ \\
\hline\hline
\end{tabular}
\end{table}

\begin{center}
\textbf{IV. SUMMARY}
\end{center}

The symmetry energy coefficients $a_{\rm sym}(A)$ for nuclei with
mass numbers $A=20 \sim 250$ have been determined from more than
2000 precisely measured nuclear masses based on the liquid drop
mass formula with the contribution of Wigner term being involved.
Taking the form of $a_{\rm sym}(A) = S_0(1+\kappa A^{-1/3})^{-1}$
we obtain the $95\%$ confidence intervals of the volume symmetry
energy coefficient $S_0 = 31.1 \pm 1.7$ MeV and the symmetry
parameter ratio $\kappa = 2.31 \pm 0.38 $ from the determined
$a_{\rm sym}(A)$.

Based on the relation between the nuclear symmetry energy at
reference density and the symmetry energy coefficients of nuclei
$e_{\rm sym}(\rho_{\rm A}) = a_{\rm sym}(A)$, we investigate the
nuclear symmetry energy at a narrow subnormal density range, $0.42
\rho_0 \leq \rho \leq 0.64 \rho_0$. Applying the reference density
for $^{208}\rm Pb$, $\rho_{208}\simeq 0.1 $ fm$^{-3}$, and the
symmetry parameter ratio $\kappa = 2.31 \pm 0.38 $, we determine
the range of $\gamma$, i.e. $\gamma = 0.7 \pm 0.1$, by inserting
the nuclear symmetry energy $e_{\rm sym}(\rho) = S_0
(\rho/\rho_0)^{\gamma}$ and the symmetry energy coefficients of
nuclei $a_{\rm sym}(A) = S_{0}(1 + \kappa A^{-1/3})^{1/\gamma}$
into the relation $e_{\rm sym}(\rho_{\rm A}) = a_{\rm sym}(A)$.
The obtained range of $\gamma$ in this way is independent on the
nuclear symmetry energy coefficient $S_0$. Simultaneously, we
deduce the mass dependence of the reference density $\rho_A$,
which explicitly depends on $\kappa$ and $\gamma$. Finally, the
range of the slope parameter $L$ of nuclear symmetry energy at
normal density is determined to be $53 \lesssim L \lesssim 79 \rm
$ MeV, based on the extracted value of $S_0$ and the extrapolated
nuclear symmetry energies at densities around the normal density.
The constraint on the nuclear symmetry energy at subnormal density
provided in this work is in good agreement with the results from
other recent analyses.

\begin{acknowledgments}
    This work was supported by the National Natural Science Foundation of China under
    Grand Nos. 10875031, 10847004, 10979023,11025524, the Doctoral Station Foundation of Ministry of Education of China under Grant No.200800270017,
    and the National Basic Research Program of China under Grant Nos. 2007CB209900, 2010CB832903.
\end{acknowledgments}

\end{document}